\documentclass[]{pasj01}

\makeatletter
\let\PASJ@origcaption\caption
\let\PASJ@origmakecaption\@makecaption
\makeatother

\let\leftroot\undefined
\let\uproot\undefined

\usepackage{xcolor}
\usepackage{amsmath}
\usepackage[round,semicolon,authoryear]{natbib}
\usepackage[implicit=False]{hyperref}
\usepackage{url}

\usepackage{listings}

\definecolor{codegreen}{rgb}{0,0.6,0}
\definecolor{codegray}{rgb}{0.5,0.5,0.5}
\definecolor{codepurple}{rgb}{0.58,0,0.82}
\definecolor{backcolour}{rgb}{0.95,0.95,0.92}

\lstdefinestyle{mystyle}{
    backgroundcolor=\color{backcolour},   
    commentstyle=\color{codegreen},
    keywordstyle=\color{magenta},
    numberstyle=\tiny\color{codegray},
    stringstyle=\color{codepurple},
    basicstyle=\ttfamily\small,
    breakatwhitespace=false,         
    breaklines=true,                 % Auto-wrap long lines
    captionpos=b,                    % Put caption at the bottom
    keepspaces=true,                 
    numbers=left,                    % Line numbers on the left
    numbersep=5pt,                  
    showspaces=false,                
    showstringspaces=false,
    showtabs=false,                  
    tabsize=2
}

\newcommand{\bx}{{\bf x}}

\Received{$\langle$reception date$\rangle$}
\Accepted{$\langle$acception date$\rangle$}
\Published{$\langle$publication date$\rangle$}
\begin{document}

\title{PSFSim: PhySics First Simulations for the Point Spread Function of the Roman Space Telescope}
\author{Nihar \textsc{Dalal} \altaffilmark{1,2}, Chris \textsc{Hirata}\altaffilmark{1,2,3}, Charuhas \textsc{Shiveshwarkar}\altaffilmark{1,2}, Anthony \textsc{Harbo Torres}\altaffilmark{1,2}, Chun-Hao \textsc{To}\altaffilmark{4,5,6}, Elise \textsc{Moore}\altaffilmark{1,2,3},
Katherine \textsc{Laliotis}\altaffilmark{1,2}, David \textsc{Kuhtenia} \altaffilmark{1,2}, Emily \textsc{Macbeth} \altaffilmark{1,2,3,7} for the Roman HLIS PIT}
\altaffiltext{1}{Center for Cosmology and Astroparticle Physics, The Ohio State University, 191 West Woodruff Avenue, Columbus, OH 43210, USA}
\altaffiltext{2}{Department of Physics, The Ohio State University, 191 West Woodruff Avenue, Columbus, OH 43210, USA}
\altaffiltext{3}{Department of Astronomy, The Ohio State University, 140 West 18th Avenue, Columbus, OH 43210, USA}
\altaffiltext{4}{Astronomy and Astrophysics Department, The University of Chicago, Chicago, IL 60637, USA}
\altaffiltext{5}{Kavli Institute for Cosmological Physics, University of Chicago, Chicago, IL 60637, USA}
\altaffiltext{6}{NSF-Simons AI Institute for the Sky (SkAI),172 E. Chestnut St., Chicago, IL 60611, USA}
\altaffiltext{7}{Steward Observatory \& Department of Astronomy, University of Arizona, 933 North Cherry Avenue, Tucson, AZ 85721, USA}
\email{dalal.64@osu.edu}


\maketitle

\begin{abstract}
We present a new tool for generating simulated point spread functions (PSFs) in the Wide Field Instrument (WFI) of the Nancy Grace Roman Space Telescope. The simulation tool is physics forward, and performs a raytrace of the optical system of the instrument to compute the PSF, while also accounting for various detector effects. The tool also has the capability to add perturbations to the optical system and detectors. We explore several applications of the tool that may arise during the commissioning of the telescope, including optical ghosts, measuring the optical aberrations from diffraction spikes, and studying the effect of mirror polarization. 
\end{abstract}

\section{Overview}

The Nancy Grace Roman Space Telescope launched on August 30th of 2026 and will conduct several community focused astrophysics surveys, including a microlensing survey for exoplanets within the Galactic bulge, a time-domain search for high-redshift supernovae, and a wide area galaxy survey for weak lensing and galaxy clustering science \citep{2019arXiv190205569A, 2025arXiv250510574O}.

The 2.4 m telescope is diffraction limited, and will image the sky in several photometric bands, with a large field-of-view of $0.281$ sq. deg. The Wide Field Instrument (WFI) focal plane consists of 18 Sensor Chip Assembly (SCA) hybrid-CMOS detectors developed by Teledyne. Each SCA consists of 4088 by 4088 pixels, and each pixel corresponds to approximately $0.11$ arcsecond on the sky, with a physical size of 10 $\mu$m \citep{WFI}. 

The point spread function (PSF) plays an important role in the characterization of all telescope images, and therefore has important and wide-reaching consequences for the science that the community can perform with the data. In the context of weak lensing, which requires precise measurements of the shapes of galaxies, the point spread function needs to be particularly well characterized to accurately measure shapes after deconvolving from the image \citep{Mandelbaum, 2021MNRAS.501.1282J, Schutt2025}. For high-redshift supernovae, an accurate PSF model is critical for accurate photometry and subtracting the contribution from a host galaxy \citep{Astier}. In the crowded microlensing fields, the PSF will impact the blending of host stars \citep{2019ApJS..241....3P}. Physical optics models of the PSF were used extensively for the Hubble Space Telescope \citep{1993ASPC...52..536K, 2011SPIE.8127E..0JK} and James Webb Space Telescope \citep{2012SPIE.8442E..3DP,2014SPIE.9143E..3XP}, for many of the concept studies that led to the Roman mission \citep{2005SPIE.5899...27S, 2008PASP..120.1307M}, and for other missions aiming for precise imaging \citep{2020MNRAS.496.5017G}.

The PSF for a space telescope is primarily determined by the optics of the instrument itself, allowing for it to be accurately and realistically characterized by simulations and optical models. Of particular importance is capturing the spatial and chromatic variation of the PSF over the large focal plane of the WFI \citep{Berlfein2025}. This motivates our package PSFSim\footnote{\url{https://github.com/Roman-HLIS-Cosmology-PIT/PSFSim}}, a physics first, ray-traced suite of simulations of the Roman PSF. Although the code was written primarily to stress test and validate the PSF modelling that will be used by the Roman High Latitude Image Survey Project Infrastructure Team (HLIS PIT), the simulation tool is fully public and of broad use to the community. Our package does not simulate the point spread function of the coronagraph, which requires a more complex optical model beyond the scope of the current work. The current version also supports only the WFI imaging modes (although it could be extended to the spectroscopic modes in the future).

In this paper, we detail the development of PSFSim, introduce the API, and compare our software to existing tools. In section \ref{sec:OpticalModel}, we describe the optical modelling that PSFSim uses to draw the point spread function. In section \ref{sec:DetectorEffects}, we describe the detector effects that PSFSim takes into account. In section \ref{sec:API}, we introduce the API for the package, and provide examples on how to draw PSFs. In section \ref{sec:Comparison}, we show that PSFSim is robust by comparing it to other PSF products produced for Roman. We describe additional applications and features of PSFSim in \ref{sec:Additional}.  

\section{Optical Model}
\label{sec:OpticalModel}

\subsection{Ray tracing}

The optical component of the point spread function is largely characterized by a ray tracing module within PSFSim. The code initializes a ray bundle at the aperture of the telescope, and follows it as it intersects, reflects, and refracts with the various optical elements, including mirrors, filters, and baffles until it reaches the WFI. The baseline positions and alignments of the optical elements are provided by the \textit{Roman} project\footnote{RST-SYS-SPEC-0055, Revision E}, with the ability to add perturbations as described in subsection \ref{sec:Perturbations}. Figure \ref{fig:coordinates} displays the coordinate convention that is used for the raytracing. 

\begin{figure}
    \centering
    \includegraphics[width=\linewidth]{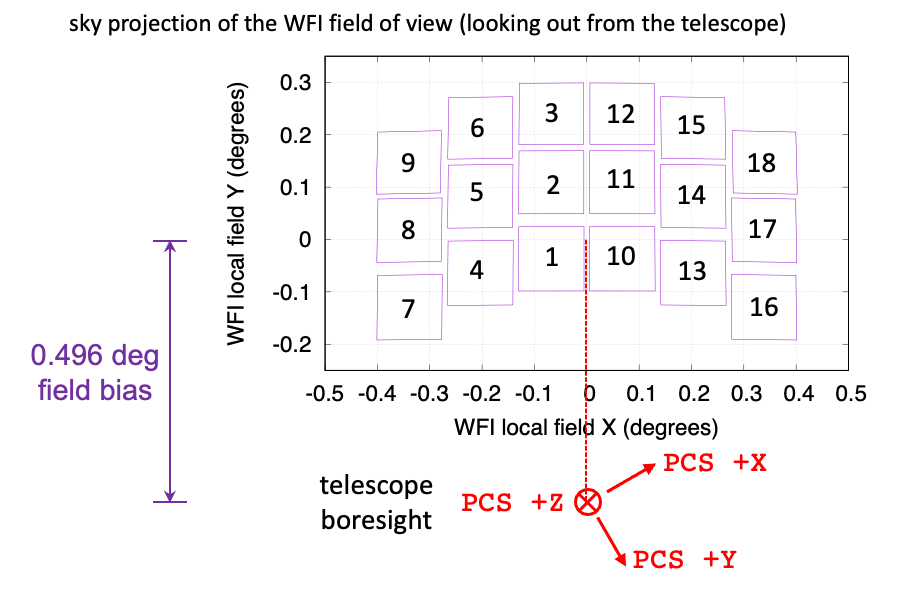}
    \caption{The coordinate definition for the local field coordinates used to initialize the ray bundle. The diagram is drawn looking at the sky.}
    \label{fig:wficoords}
\end{figure}

\begin{figure}
  \begin{center}
    \includegraphics[width=\linewidth]{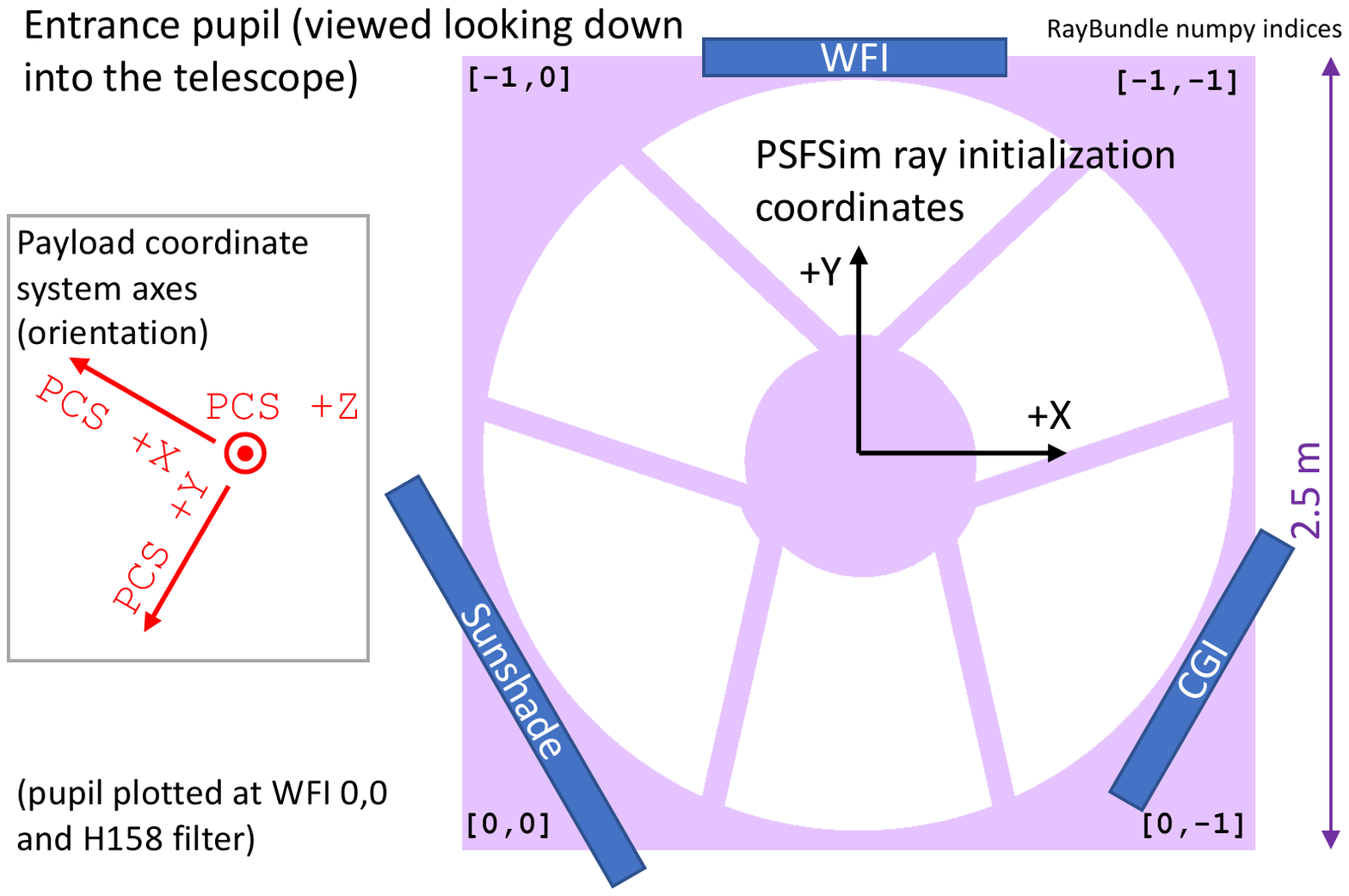}
  \end{center}
  \caption{Coordinate description of the initial ray bundle used in ray trace}
  \label{fig:coordinates}
\end{figure}

\begin{figure*}
    \centering
    \includegraphics[width=6in]{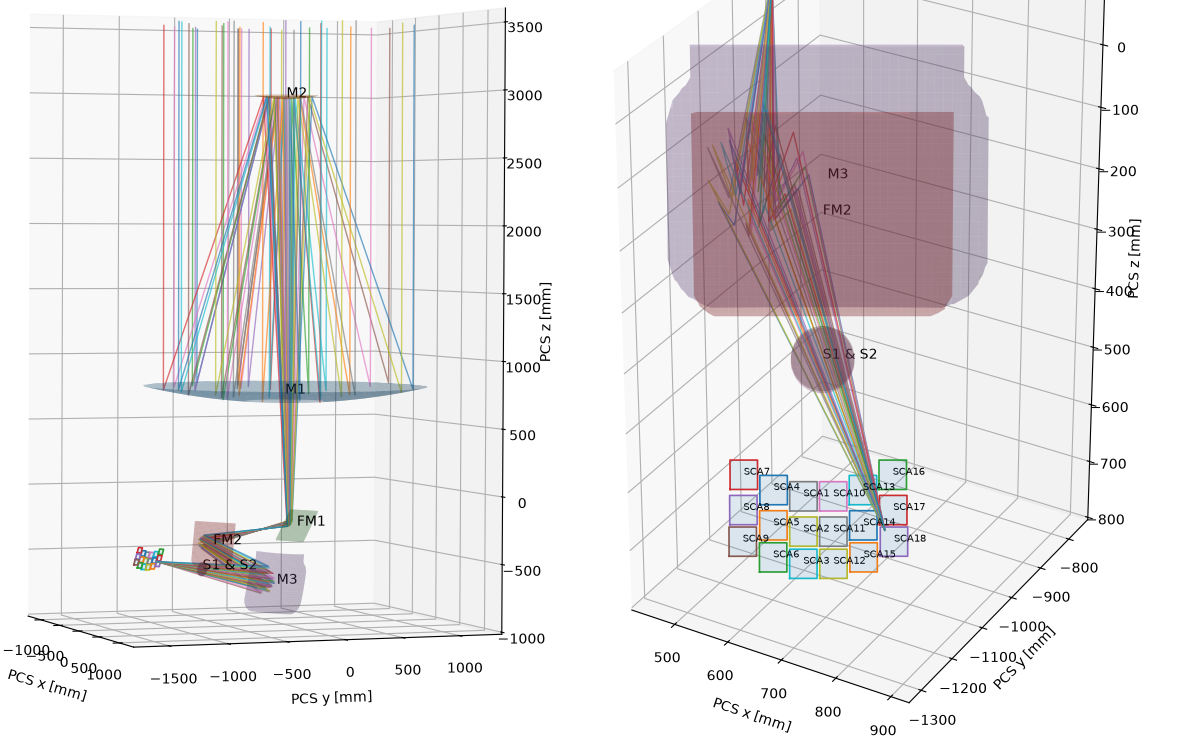}
    \caption{Full view (left panel) of the optical path using PSFSim, and zoom-in on the FPA, filter surface 1 and 2, the second fold mirror and the tertiary mirror.}
    \label{fig:optsystem}
\end{figure*}

\begin{figure*}
    \centering
    \includegraphics[width=0.44\linewidth]{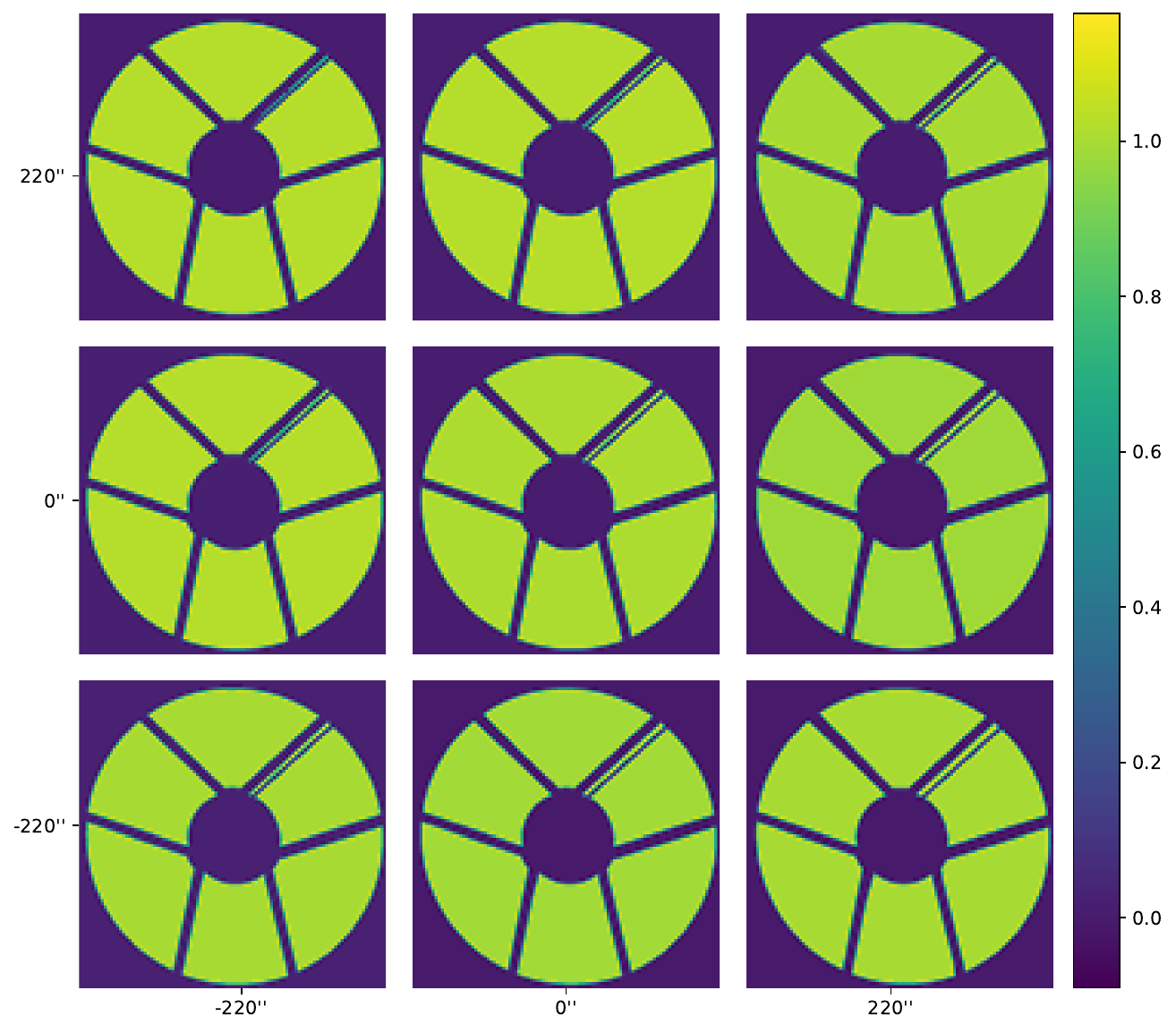}
    \includegraphics[width=0.5\linewidth]{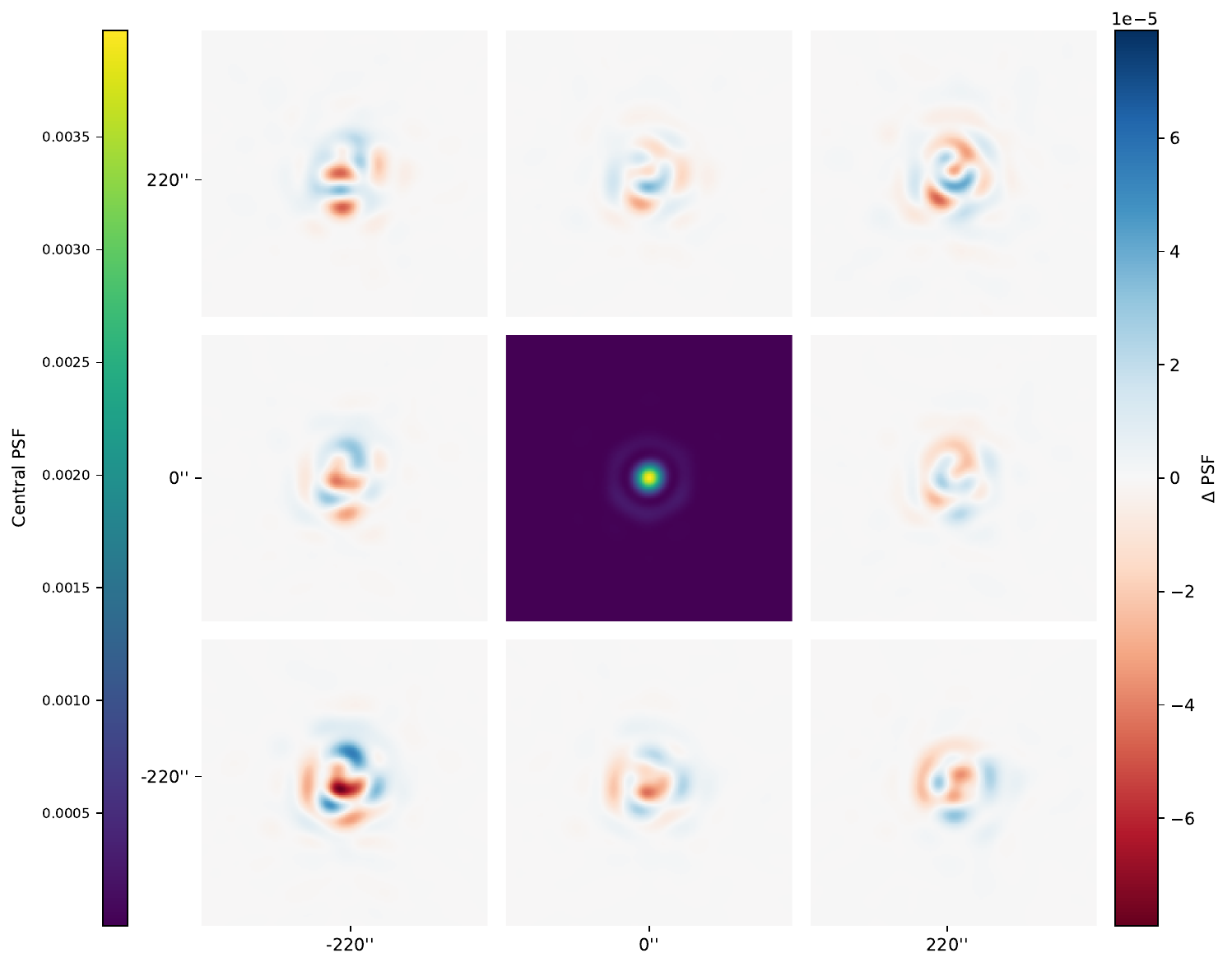}
    \caption{Left: Spatial variation in the PSFSim computed pupil mask in SCA07. We note the secondary shadow of the top right strut changes in width as you move across the chip. Right: Spatial variation in the SCA07 polychromatic PSF. In the centre panel, we have the baseline PSF, and in each of the corners we have the difference in the PSF computed at that location with respect to the centre.}
    \label{fig:spatial_variation}
\end{figure*}

After the ray bundle is initialized, it is first obstructed by the six support tubes for the secondary mirror, which form the twelve diffraction spikes in the point spread function, and generate the particular shape of the pupil mask, as shown in the right panel of Figure \ref{fig:coordinates}. The ray bundle is then further masked by baffles for the secondary mirror, and later the baffles for the primary mirror, before finally reflecting off the primary mirror. The light then reflects from the secondary mirror, before passing through the hole in the primary mirror. The light then hits the fold mirror, and is subsequently masked by the entrance aperture plate. Finally, the light hits the second fold mirror, and the tertiary mirror before exiting the `pupil mask' $\mathcal{M}(u,v)$. A schematic diagram of the optical system is provided by the project in Figure \ref{fig:optsystem}, and the resultant pupil as it varies spatially is presented in Figure \ref{fig:spatial_variation}.

We begin the ray tracing with a bundle of parallel rays
at angles $r$ and $\phi$, where $x_{\rm an} = r\cos\phi$ and $y_{\rm an} = r\sin\phi$.
%\begin{align}
%    \phi  &= \tan^{-1} (y_{\rm{an}}/x_{\rm{an}})\\
%    r &= \sqrt{x_{\rm{an}}^2 +y_{\rm{an}}^2}
%\end{align}
Here $r$ is in radians, and $x_{\rm{an}}, y_{\rm{an}}$ refer to the local WFI coordinate system (see Fig \ref{fig:wficoords}). 
We also track the polarization of the ray bundle, separately propagating the complex vector amplitudes for the 2 polarizations. The incoming electric field in the horizontal (H) and vertical (V) polarizations ${\bf E}^{\rm H} , {\bf E}^{\rm V} $ directions has components:
\begin{equation}
    {\bf E}^{\rm H} = \begin{pmatrix}
    \sin^2 \phi + \cos r\cos^2\phi\\
    (\cos r-1) \sin \phi \cos\phi\\
    -\sin r \cos \phi
    \end{pmatrix}
\end{equation}
and
\begin{equation}
{\bf E}^{\rm V} = \begin{pmatrix}
    (\cos r-1) \sin \phi \cos\phi\\
    \cos^2 \phi + \cos r\sin^2\phi\\
    -\sin r \sin \phi
    \end{pmatrix}.
\end{equation}

This electric field is then rotated into the Payload Coordinate System (PCS) by removing the field bias (0.496 degrees from the center of the WFI field angle system to the telescope boresight) and doing the 150 degree rotation from the WFI field angle +Y direction to PCS +Y. The electric fields ${\bf E}^{\rm H}$ and ${\bf E}^{\rm V}$ are propagated independently using the S- and P-polarized reflection and transmission coefficients at each surface until they reach the detector. There, both the position and propagation direction vectors and the electric fields are rotated to FPA coordinates. Each ${\bf E}$ remains perpendicular to the light ray on the focal plane. Specifically, in terms of ${\bf p} = (u, v, \sqrt{1-u^2-v^2})$ we have the condition
\begin{equation}
    {\bf E}^{\rm H(V)} \cdot \bf p = 0.
\end{equation}
We can also compute the magnetic field as 
\begin{equation}
    c{\bf B}^{\rm H(V)} = {\bf p} \times {\bf E}^{\rm H(V)}.
\end{equation}

\subsection{Optical aberrations}

The baseline optical aberrations of the telescope are characterized by the Zernike coefficients $C_j(x,y)$. PSFSim uses design data from the WIM Cycle 9 to compute the phase difference at a given position $\bx_{\rm{out}}$. The Cycle 9 data provide 22 Zernike coefficients measured at 5 different points on the detector (centre and four corners), and measured at 18 different wavelengths spanning the range from $480 \, \rm{nm}$ to $2000 \, \rm{nm}$. For the updated Cycle 10 reference data (and planned updates in the future), we instead use the perturbation model described in Section \ref{sec:Perturbations} below. The Zernike coefficients are in the NOLL form with index $j$, which can be obtained from the regular pair of Zernike indices $m,n$ via the following:
\begin{equation}
    j=\frac{n(n+1)}{2} + |m| + \begin{cases}
        1 & \text{ if } (-1)^{\lfloor n/2\rfloor} m \leq 0\\
        0 &  \text{ otherwise}
    \end{cases}
\end{equation}

% \begin{equation}
%     j = \frac{n(n+1)}{2} + |m|+ \begin{cases}
%     0, \, m >0 \, \land \, n \equiv \{0,1\} \pmod 4\\ 
%     0, \, m <0 \, \land \, n \equiv \{2,3\} \pmod 4\\ 
%     1, \, m \geq 0 \, \land \, n \equiv \{2,3\} \pmod 4\\ 
%     1, \, m \leq 0 \, \land \, n \equiv \{0,1\} \pmod 4\\ 
%     \end{cases}
% \end{equation}
The even Zernike polynomials are defined as always, in terms of a radial coordinate $\rho$ and an angle $\alpha$ as 
\begin{equation}
    Z_{n}^m(\rho, \alpha) = 
        R_n^m(\rho) \cos(m\alpha)
\end{equation}
where the radial Zernike polynomial is
\begin{equation}
    R_n^m = \sum_{k=0}^{\frac{n-m}{2}} \frac{(-1)^k (n-k)!}{k!\left(\frac{n+m}{2}-k \right)!\left(\frac{n-m}{2}-k \right)!} \rho^{n-2k}.
\end{equation}
The odd Zernike polynomials are given instead by 
\begin{equation}
    Z_{n}^{-m}(\rho, \alpha) = 
        R_n^m(\rho) \sin(m\alpha)
\end{equation}
We use linear grid interpolation to solve for the path differences at any point on the detector, and once again linearly interpolate over the desired wavelength. The resultant baseline optical path difference is therefore obtained at a desired position on the detector $x,y$ as a function of $u,v$ as 
\begin{equation}
    \Delta \phi(u,v) = \sum_{j=1}^{22} N_jC_j(x,y) Z_j\Big(\sqrt{u^2+v^2}, \tan^{-1}\frac vu\Big),
\end{equation}
using the Noll normalization conventions ($N_j = \sqrt{n+1}$ if $m=0$ and $\sqrt{2(n+1)}$ if $m\neq 0$).

% \begin{figure}
%     \centering
%     \includegraphics[width=\linewidth]{Figures/focal_plane_all_psfs.pdf}
%     \caption{The variation of the polychromatic Optical PSF in H band as computed by PSFSim over the focal plane. Each PSF is computed at the center of the SCA, and normalized to the same intensity.}
%     \label{fig:focal_plane_all_psf}
% \end{figure}

\begin{figure*}
    \centering
    \includegraphics[trim={4cm 1cm 6cm 1cm}, clip=true, width=\linewidth]{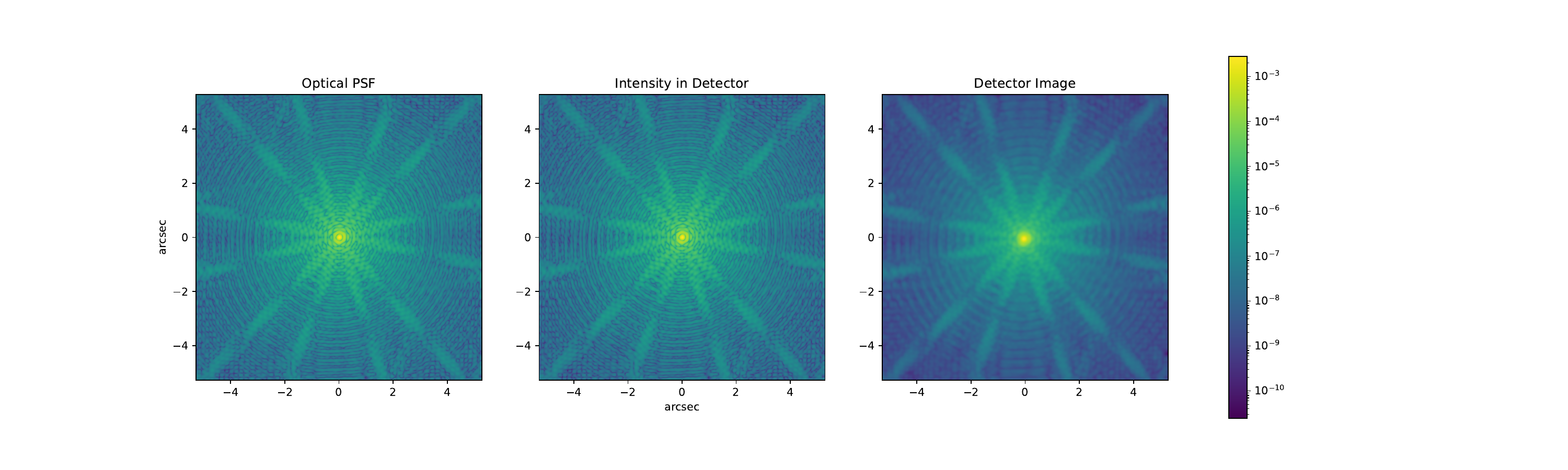}
    \caption{Stages of the monochromatic PSF. Left: the purely optical PSF that is the result of ray tracing through the telescope, and with the baseline optical aberrations. Middle: the intensity in the detector as the electric field is propagated through the detector layers. Right: the image as seen on the detector, after the intensity has been convolved with the Modulation Transfer Function. }
    \label{fig:monochrom}
\end{figure*}

\subsection{Perturbations to the Optical Model}
\label{sec:Perturbations}

The ``as-built'' telescope is not exactly the same as the design specification. Furthermore, certain optical elements can intentionally be perturbed to improve the optical performance of the telescope or for wavefront sensing. PSFSim treats this by perturbing the optical elements to match a given distortion map and Zernike coefficients. These currently come from the Roman Project's table of predicted coefficients, but will be updated based on the in-flight measurement.

It is not possible to derive the full set of perturbations in this way, because there are some degeneracies between the motions of various optical elements --- for example, one could move a fold flat mirror, and move the downstream components to the position of the new image, without changing the wavefront. Therefore, we allow only a subset of perturbations, and note that while fitting these perturbations should lead to a good model of the PSF, this does not mean that the model perturbation is the actual perturbation in the hardware.

%Therefore, in addition to the baseline optical system and path differences, we add in the ability to shift around certain elements to alter the path differences via user specified perturbations.

%\nd{More specific description of which elements are actually perturbed from Chris?}

The perturbations are performed in two steps. First is the offset of the element positions, which is least-squares fit to the distortion map. This has 9 components: the 6 degrees of freedom of the focal plane; the Secondary Mirror z-position; and the 2 translations of the field angles (these are technically degenerate with the WFI pointing, but if we want the perturbed model to match a distortion map then we need to match the field center in the model to the field center in the distortion map). These 9 degrees of freedom allow for the translation (2), rotation (1), magnification (1) and shear (2) of the field angle to focal plane transformation; as well as the center of the distortion map (2) and the long-axis ``rectangle to trapezoid'' mode of the distortion (1).

The second step is a set of perturbations to the optical surfaces. These are described by a set of basis modes and implemented by a path length excess added at each reflection and refraction. The current set of perturbations used for the Cycle 10 model are:
\begin{itemize}
    \item Primary mirror: Zernike basis, orders $2\le n\le 6$.
    \item Tertiary mirror: 2D Legendre basis, orders $\ell_x\le 5$, $\ell_y\le 4$, $\ell_x+\ell_y \ge 3$, where $\ell_x$ and $\ell_y$ correspond in a consistent way to the orders of the components $x$ and $y$ of the polynomials.
    \item Focal plane: 2D Legendre basis, $\ell_x + \ell_y\le 1$ (i.e., piston + tip/tilt) for each of the 18 SCAs.
    \item Filter: surface S1, Zernike basis, orders $2\le n\le 6$. (There is a different set of coefficients for each filter.)
\end{itemize}
This leads to 128 perturbation coefficients in a single filter, or 303 taking into account that 25 of these parameters are re-fit for each of the 8 filters. Note that subsets of the possible physical perturbations were taken to avoid degeneracies. The resulting residuals from the Cycle 10 data was 12 nm (root-sum-squared over the Zernike coefficients, and with a root-mean-square taken over the field positions in the Cycle 10 data).

\subsection{Optical Point Spread Function}
\label{sec:psf}

The profile of the ``optical" point spread function in real space can be computed as 
\begin{equation}
    I_{\rm{PSF}}(x,y) = \frac{1}{2} \big[ I^{\rm H}_{\rm{PSF}} (x,y) + I^{\rm V}_{\rm{PSF}} (x,y)\big], 
\end{equation}
where the horizontally (vertically) polarized PSF is further given by the magnitude of the Poynting vector 
\begin{equation}
    I^{H(V)}_{\rm{PSF}} (x,y) = \frac{1}{2} \epsilon_0 c^2 \mathrm{Re}\big[ {\bf \tilde{E}}^{\rm H (V)}(x,y) \times {\bf \tilde{B}
    }^{\ast\rm H(V)}(x,y) \big],
\end{equation}
where ${\bf \tilde{E}}(x,y)$ and ${\bf \tilde{B}}(x,y)$ are 2D Fourier transforms of the ray traced electric and magnetic fields ${\bf E}(u,v)$ and ${\bf B}(u,v)$ as computed by the raytrace module, alongside the necessary prefactors. More explicitly, where ${\bf u} = (u,v)$ and ${\bf x} = (x,y)$, we have

\begin{eqnarray}
    \tilde{\bf E}(x,y) &=& \int d^{2}\textbf{u}\ \mathcal{M} (u,v)\cdot\det\left(\frac{\partial \textbf{x}_{\rm in}}{\partial\textbf{u}}\right)_{x,y}\nonumber\\
    &&\times\exp\left(i\frac{2\pi}{\lambda}\left(\Delta\phi(u,v)+\textbf{u}\cdot\textbf{x}\right)\right)\cdot\textbf{E}(u,v)\
\end{eqnarray}
and 
\begin{eqnarray}
    \tilde{\bf B}(x,y) &=& \int d^{2}\textbf{u}\ \mathcal{M} (u,v)\cdot\det\left(\frac{\partial \textbf{x}_{\rm in}}{\partial\textbf{u}}\right)_{x,y}\nonumber\\
    &&\times\exp\left(i\frac{2\pi}{\lambda}\left(\Delta\phi(u,v)+\textbf{u}\cdot\textbf{x}\right)\right)\cdot\textbf{B}(u,v).\
\end{eqnarray}

The optical point spread function is propagated downstream to the detector, where it is further modified by various detector effects which are described below in Section \ref{sec:DetectorEffects}. The various stages of the monochromatic PSF are also displayed in Fig \ref{fig:monochrom}. The monochromatic optical point spread function is an attribute that can be accessed as demonstrated in the codeblock below:

\begin{lstlisting}[language = python]
import numpy as np
from psfsim.psfsobject import PSFObject
import matplotlib.pyplot as plt
mypsf = PSFObject(9,0,0, wavelength = 1.0)
mypsf.get_optical_psf()
plt.imshow(mypsf.Optical_PSF)
\end{lstlisting}

\section{Detector Effects}
\label{sec:DetectorEffects}
\subsection{Interference Layers}

The Wide Field Instrument focal plane is populated with 18 H4RG-10 detectors. Each H4RG-10 detector is an HgCdTe (mercatel) 4096 pixel by 4096 pixel photodiode array with a 10 micron pixel pitch \citep{Mosby2020, Mosby2025}. The photon sensitive HgCdTe layer is bonded to a silicon read out circuit, which is further connected to a mechanical mount with electrical connections. The mole fraction of cadmium has been tuned to $0.445$ \citep{Mosby2020} to achieve the desired cutoff wavelength at 2.5 microns. The HgCdTe layer also improves the quantum efficiency and the count rate non-linearity of the detectors. 

The detector also includes a proprietary anti-reflection (AR) coating to ensure that the in-band light is transmitted at high efficiency. The AR coating is the final optical surface in the WFI, and it effectively functions as an interference layer. In the current version of PSFSim, the AR coating is an ideal 3-layer model, with geometrically spaced indices of refraction $n$ and equal optical thicknesses $nt$ (where $t$ is the physical thickness). This covers the HgCdTe absorber with a complex index of refraction \citep{Mercadtel99}. A very thin absorbing layer was also included just above the HgCdTe. (PSFSim does {\em not} have the physically correct model for the AR coating/HgCdTe interface since it contains no proprietary data; but what is important here is to get the correct susceptance of the absorbing layer to match the quantum efficiency curve. This will be updated with an empirical model in flight.)

To propagate the PSF further into the detector, we need to decompose the incoming electromagnetic wave into transverse electric (TE) and transverse magnetic (TM) modes. For this decomposition, we construct a local coordinate system centered at the point of incidence (POI), with the z-axis of this local coordinate system pointing inside the detector and (for $u^2 + v^2\neq 0$):
\begin{align}
    \hat{y}_{\rm POI} &= \hat{u} = \frac{(u,v,0)_{\rm FPA}}{\sqrt{u^2+v^2}} = (\cos \phi_{\rm inc}, \sin \phi_{\rm inc}, 0)_{\rm FPA}, \nonumber\\
    \hat{z}_{\rm POI} &= \hat{z}_{\rm FPA},\ {\rm and}\nonumber\\
    \hat{x}_{\rm POI} &= \hat{y}_{\rm POI}\times\hat{z}_{\rm POI}\ .
\end{align}
For normal incidence (where $u^2 + v^2 = 0$), we choose the x- and y-axes of the POI coordinate system along the x and y axes of the FPA coordinates. 
% %Chris: I don't think we need to write this part explicitly.
%\begin{align}
%    \hat{x}_{\rm POI} &= \hat{x}\\
%    \hat{y}_{\rm POI} &= \hat{y}\\ 
%    \hat{z}_{\rm POI} &= \hat{z}\ .
%\end{align}
Note that the POI frame (defined by the basis vectors $\hat{x}_{\rm POI}, \hat{y}_{\rm POI}, \hat{z}_{\rm POI}$) is defined such that the x-axis of this frame is perpendicular to the plane of incidence. Note that for normal incidence, the POI frame and the FPA frame are identical.

With the above choice of POI coordinates we can define the local TE and TM polarisation directions $\hat{\epsilon}_{TE}$ and $\hat{\epsilon}_{TM}$ of the electric field as follows 
\begin{align}
    \hat{\epsilon}_{\rm{TE}} &= \hat{x}_{\rm POI} \nonumber \\
    &= \frac{(v, -u, 0)_{\rm{FPA}}}{\sqrt{u^2+v^2}} 
    \nonumber \\
    &= (\sin \phi_{\rm{inc}}, -\cos \phi_{\rm{inc}}, 0)_{\rm FPA}
    ~{\rm and}\nonumber \\
    \hat{\epsilon}_{\rm{TM}} &= \hat{x}_{\rm POI}\times \hat{k} \nonumber \\ &= \frac{(-u\sqrt{1-u^2-v^2}, -v\sqrt{1-u^2-v^2}, u^2+v^2)_{\rm{FPA}}}{\sqrt{u^2+v^2}}\ . 
    %\\
    %&= \sqrt{1-u^2-v^2} \left(-\cos\phi_{\rm{inc}}, -\sin\phi_{\rm{inc}}, \sqrt{\frac{u^2+v^2}{1-u^2-v^2}}\right)\ .
    % C.H.: I don't think we need to write the last version in the text.
\end{align}
Note that for normal incidence, we have
%have $\phi_{\rm{inc}} = \frac{\pi}{2}$,
$\hat{\epsilon}_{\rm{TE}} = (1,0,0)_{\rm FPA}$ and $\hat{\epsilon}_{\rm{TM}} = (0,-1,0)_{\rm FPA}$.

With the above choice of local coordinates, we can obtain amplitudes $\mathcal{A}_{TE}$ and $\mathcal{A}_{TM}$ of the TE and TM modes. Note that $\mathcal{A}_{TE}$ ($\mathcal{A}_{TM}$) is the amplitude of the electric (magnetic) field along the $\hat{x}_{\rm POI}$ direction: 
\begin{equation}
    \mathcal{A}_{TE} = \textbf{E}_{0}\cdot\hat{x}_{\rm POI}
    ~~{\rm and}~~
    \mathcal{A}_{TM} = \left(\hat{\textbf{k}}\times\textbf{E}_{0}\right)\cdot\hat{x}_{\rm POI}\ .
\end{equation}

The light incident on the detector passes through an interference filter tri-layer of some thickness $\Delta = t_{1}+t_2 + t_3$. For future calculations, it's helpful to introduce here a coordinate system centered at the pixel center (i.e. at the point with position vector $\textbf{x}_{\rm out} + \textbf{s} + \Delta\hat{\textbf{z}}$ in the FPA coordinates). A position vector of $\textbf{r}_{p} = (x_p,y_p,z_p\geq 0)$ in this coordinate system denotes a point inside the HgCdTe layer below the pixel.  
Following the treatment of~\citep{Born:1999ory} about electromagnetic wave propagation in stratified media, 
we express the electric field  at a point $\textbf{r}_p$ as follows :
\begin{eqnarray}
    \textbf{E}\cdot\hat{x}_{\rm POI} &=& \mathcal{A}_{TE}\mathcal{T}_{TE}\exp\left(i\beta z_{p} + i\frac{2\pi}{\lambda}\textbf{u}\cdot\textbf{r}_p\right), \nonumber \\
    \textbf{E}\cdot\hat{y}_{\rm POI} &=& \frac{-\beta}{k_{0}\epsilon}\mathcal{A}_{TM}\mathcal{T}_{TM}\exp\left(i\beta z_{p} + i\frac{2\pi}{\lambda}\textbf{u}\cdot\textbf{r}_p\right), ~{\rm and}\nonumber \\
    \textbf{E}\cdot\hat{z}_{\rm POI} &=& \frac{|\textbf{u}|}{\epsilon}\mathcal{A}_{TM}\mathcal{T}_{TM}\exp\left(i\beta z_{p} + i\frac{2\pi}{\lambda}\textbf{u}\cdot\textbf{r}_p\right)\ ,
\end{eqnarray}
where $\mathcal{T}_{TE}$ and $\mathcal{T}_{TM}$ are the $\textbf{u}$-dependent transmission coefficients of the tri-layer for the TE and TM modes respectively, $k_{0} = 2\pi/\lambda$, $\epsilon$ is the wavelength-dependent permittivity of HgCdTe and $\beta$ is the root of the equation $\beta^2 = k_{0}^{2}(n_{\rm HgCdTe}^{2}-|\textbf{u}|^{2})$ with positive imaginary part. We then rotate the electric field components in the POI frame into components along the FPA basis. For non-normal incidence\footnote{For normal incidence, the $\left(\hat{x}_{\rm POI}, \hat{y}_{\rm POI},\hat{z}_{\rm POI}\right)$ frame is identical to the FPA basis by construction.}, we then rotate the electric field components into components along the POI basis, whereby we have 
\begin{eqnarray}
    E_{x} &=& \frac{u_{y}}{u} \textbf{E}\cdot\hat{x}_{\rm POI}\ +\ \frac{u_{x}}{u}\textbf{E}\cdot\hat{y}_{\rm POI}, \nonumber \\
    E_{y} &=& -\frac{u_{x}}{u}\textbf{E}\cdot\hat{x}_{\rm POI}\ +\ \frac{u_{y}}{u}\textbf{E}\cdot\hat{y}_{\rm POI}, ~{\rm and} \nonumber\\
    E_{z} &=& \textbf{E}\cdot\hat{z}_{\rm POI}\ .
\end{eqnarray}

This eventually returns an intensity in the detector
\begin{equation}
    I_{\rm detector} \propto|E|^2_{\rm detector}\ .
\end{equation}

\subsection{Modulation Transfer Function \& Charge Diffusion}
The modulation transfer function (MTF) of the detector describes its spatial frequency response to a sine wave fluctuation in the sky image and is a combination of the ideal pixel top hat response, charge diffusion, and interpixel capacitance (IPC). We use the MTF as measured and modelled by \cite{Macbeth2026} using data from the laser speckle tests in the Detector Characterization Lab. 

The MTF is best modelled with a hyperbolic secant profile, which is well approximated as the sum of three gaussians. Specifically, we have 
\begin{equation}
    {\rm MTF}_{\rm diffusion} = \mathrm{sech}(2\pi\sigma u) \approx \sum_{i=1}^{3} w_i e^{-2\pi^2 (c_i\sigma)^2 u^2}
\end{equation}
where the scaling of the widths are $c_i = \{0.4522, 0.8050, 1.4329 \}$, the relative weights $w_i = \{0.17519, 0.53146, 0.29335\}$, and the normalization is $\sigma = 0.3279$ \citep{Macbeth2026, OU24}. We also note that \citet{Macbeth2026} finds little to negligible wavelength dependence in the MTF as measured by the speckle tests, so we do not incorporate any wavelength dependence here. 

To get a realized detector image, we convolve the detector intensity with the modeled MTF:
\begin{equation}
    I_{\rm detector} \propto |E|^2_{\rm detector} \star {\rm MTF}_{\rm diffusion} \star {\rm MTF}_{\rm tophat}.
\end{equation}
This returns a realization of an image on the detector, accounting for both optical and detector contributions. We do {\em not} include inter-pixel capacitance or electronic cross-talk effects here: these are intended to be modeled (or, if necessary, corrected) outside of PSFSim. The object can be returned by the code for a particular wavelength and at a position as described by the codeblock below:

\begin{lstlisting}[language=python]
import numpy as np
from psfsim.psfsobject import PSFObject
import matplotlib.pyplot as plt
mypsf = PSFObject(9,0,0, wavelength = 1.0, postage_stamp_size = 32, oversamp = 10)
mypsf.get_image_from_Intensity()
plt.imshow(mypsf.detector_image)
\end{lstlisting}

\section{API}
\label{sec:API}
In this section, we briefly detail the API for using PSFSim. Extensive documentation for the code is maintained and updated on our readthedocs page \url{https://psfsim.readthedocs.io/en/latest/index.html}. In most cases, the user would interact primarily with the \verb|PolychromaticPSF| class of the \verb|psfsim.polychrom| module. The example code block below describes how to instantiate the class, as well as compute a polychromatic PSF over a flat SED in a given bandpass. 

\begin{lstlisting}[language=python]
import numpy as np
from psfsim.polychrom import PolychromaticPSF
import matplotlib.pyplot as plt
h_band_wl = np.linspace(1.4, 1.8, 10)
poly = PolychromaticPSF(9,0,0, h_band_wl, sed = None)
poly.compute_poly_psf(postage_stamp_size = 32, optical_psf_only = True, use_postage_stamp_size = 96, ovsamp = 10)
plt.imshow(poly.chromatic_psf)
\end{lstlisting}

The \verb|compute_poly_psf| method computes and sums monochromatic PSFs at the desired size given by the keyword argument \verb|use_postage_stamp_size| and oversampled by a factor given by \verb|ovsamp| argument. The sums can be weighted by an \verb|sed| argument that is specified when the \verb|PolychromaticPSF| class is instantiated. The user can choose to output an `optical only' PSF with the \verb|optical_psf_only| argument, or a PSF convolved with the modulation transfer function of the detector.

%\Chris{add description of the Science Frame API}\nd{Done below}

Given the several different coordinate systems used by the Roman project, our code also has the ability to compute PSFs in various frames. The examples above have been in the \verb|analysis| frame, where the coordinates are specified in $\mu$m with respect to the center of a given SCA. The PSF can also be computed in the \verb|science| frame and the \verb|FPA| frame, which are measured in pixels, and with the respective orientations as displayed in Figure \ref{fig:frames}. We note that STPSF is in the \verb|FPA| frame. The user can specify the frame with the \verb|frame| optional argument as below.
\begin{lstlisting}[language=python]
poly = PolychromaticPSF(9, 2044, 2044, h_band_wl, frame = 'science')
\end{lstlisting}

\begin{figure*}
    \centering
    \includegraphics[width=\linewidth]{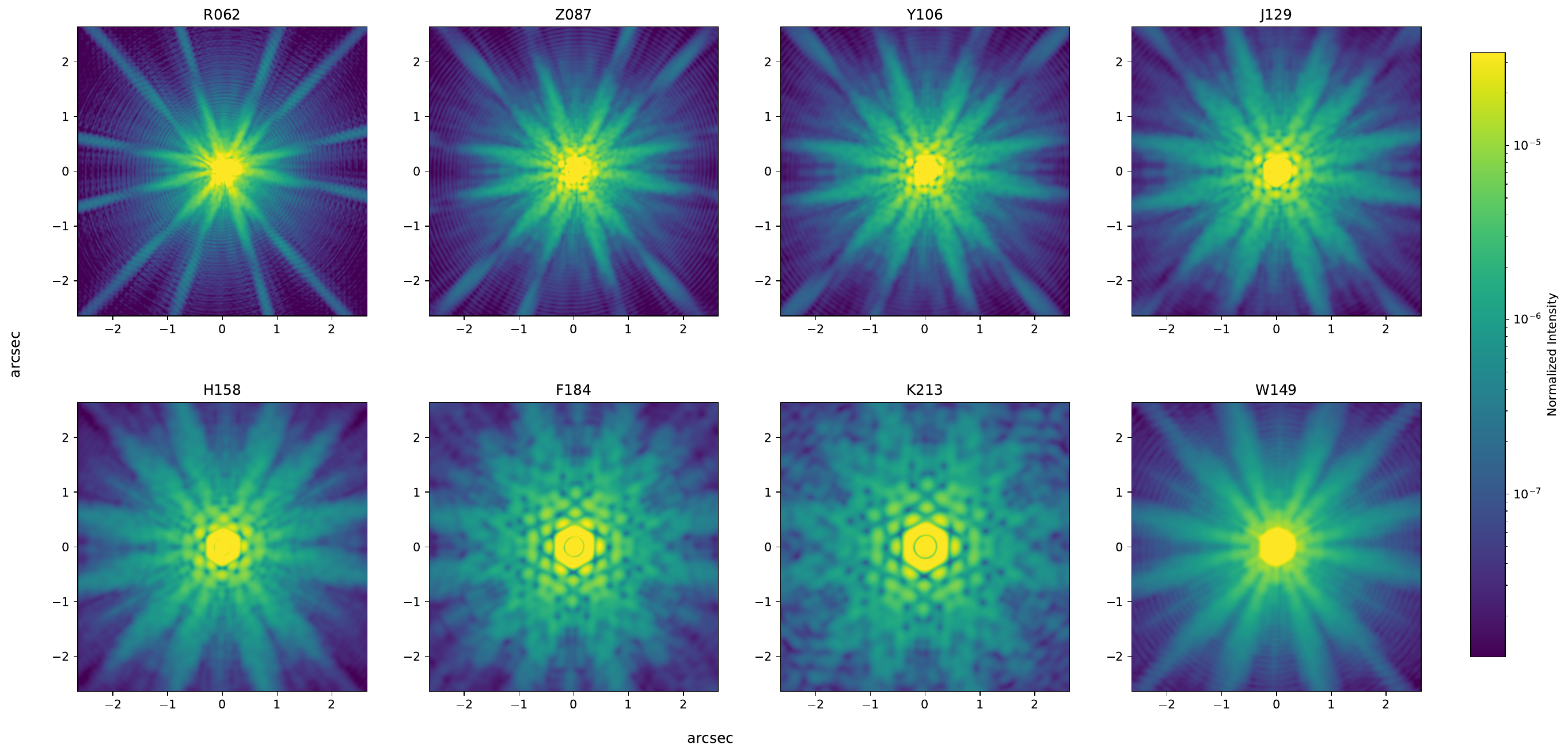}
    \caption{The Roman Optical PSF as it varies over different filters at the center of SCA09 \footnote{Note that we adopt the convention of referring to the filters with their common names. However, the official filters names are F062, F087, etc.}.}
    \label{fig:bandpass}
\end{figure*}

\begin{figure}
    \centering
    \includegraphics[width=\linewidth]{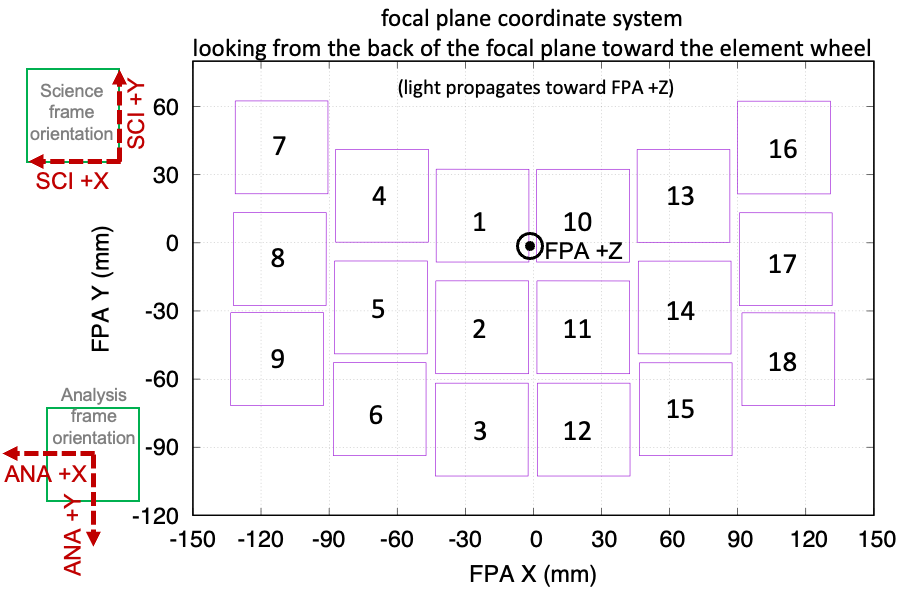}
    \caption{A description of the various frames and axes orientations used in PSFSim.}
    \label{fig:frames}
\end{figure}

\begin{figure}
    \centering
    \includegraphics[width=\linewidth]{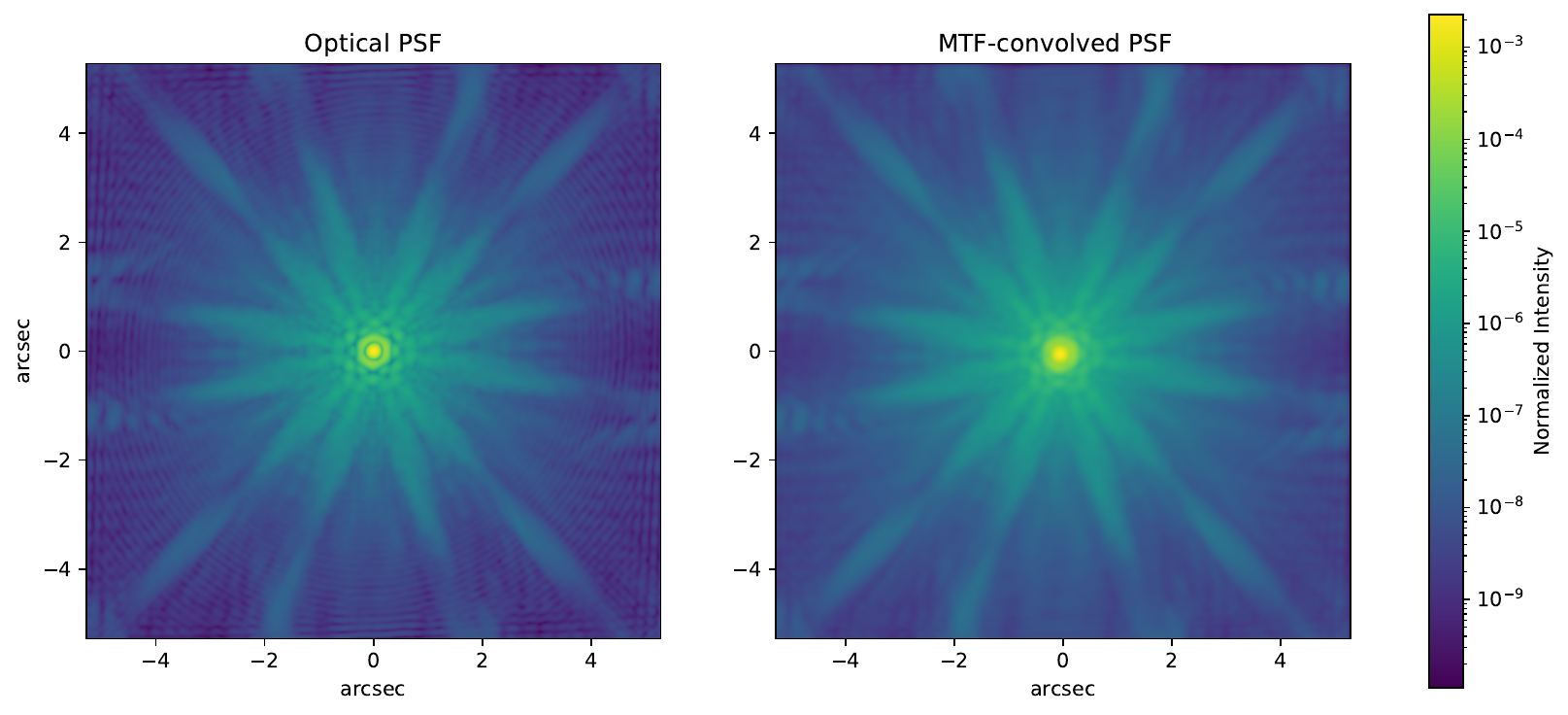}
    \caption{The two stages of the polychromatic PSF. Left: the purely optical PSF over the H band. Right: the H band PSF convolved with the Modulation Transfer Function measured by \cite{Macbeth2026}.}
    \label{fig:Polychrom}
\end{figure}

We display the change in the PSF over various bandpasses in Fig \ref{fig:bandpass}, as well as a description of the various coordinates and frames in Fig \ref{fig:frames}. Finally, in Fig \ref{fig:Polychrom}, we display the two stages of the polychromatic PSF - on the left is the purely optical polychromatic PSF, and on the right is the polychromatic PSF convolved with the modulation transfer function.  

\section{Comparisons to existing simulations}
\label{sec:Comparison}
In order to test the fidelity of our simulations, we compare the output point spread functions to those modelled for Roman by the STPSF\footnote{\url{https://stpsf.readthedocs.io/en/latest/roman.html}} package. The STPSF package is developed and maintained by the Roman Project at Space Telescope Science Institute, and relies on the Poppy\footnote{\url{https://poppy-optics.readthedocs.io/en/latest/installation.html}} optics package \citep{2012SPIE.8442E..3DP, 2014SPIE.9143E..3XP}. STPSF and Poppy have successfully been used to model the point spread function of the James Webb Space Telescope.

STPSF uses Cycle 10 reference information provided by the Roman team at Goddard Space Flight Center. The reference information available gives the field dependent aberrations in terms of Zernike polynomial coefficients from $Z_{1}$ to $Z_{45}$ for five field points on each detector (corners and center), and with coverage from $0.76 \, \mu m$ to  $2.3 \, \mu m$. STPSF interpolates the coefficients in position and wavelength space to allow users to simulate PSFs at any valid pixel position and wavelength. STPSF approximates the aberrations for an out-of-range detector position by using the nearest field point. The pupil masks used by STPSF also have field dependence that varies SCA by SCA, as well as some coarse wavelength dependence where the shorter wavelength filters use a wider pupil mask.  

\begin{figure*}
    \centering
    \includegraphics[width=\linewidth]{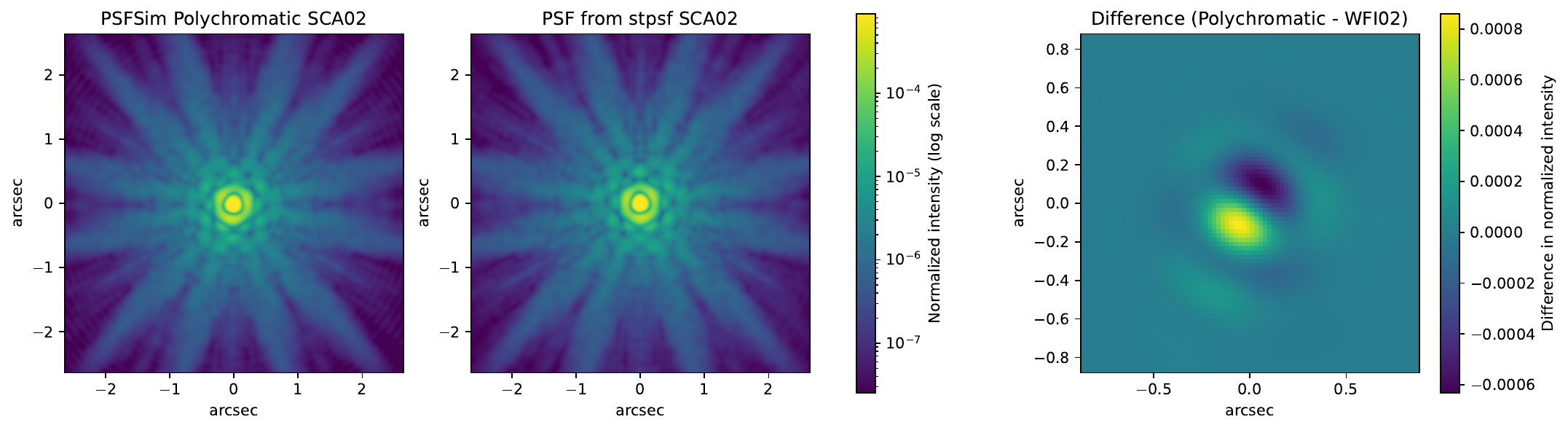}
    \caption{Comparison between PSFSim (left panel) and STPSF (middle panel), and the difference image (right panel). Both PSFs have been drawn at the center of SCA2 in the H158 band. There is some residual structure in the difference image that we attribute to a combination of implementation differences between the two packages.}
    \label{fig:stpsfvs}
\end{figure*}

In Fig \ref{fig:stpsfvs}, we compare the PSF as generated by PSFSim (left panel) to the PSF generated by STPSF (middle panel). Both PSFs are drawn at the center of SCA2 in the H158 band, and assuming Cycle 10 design data. The difference image is shown in the right panel, and displays a clear dipole pattern, and a faint annulus. We find that this difference has a field dependence, and the dipole changes orientation across the focal plane. The dipole pattern is likely a result of the difference in centering between the two codes, as STPSF and PSFSim treat the astrometry differently.  

We attribute this difference to a variety of features that differ between the two packages, most notably the treatment of refraction, as well as the models of charge diffusion. We find that the difference bears similar patterns to the decentering and defocus due to refraction that is presented in Figure 5 of \cite{Berlfein2026}.

Nonetheless, the general agreement between the two packages is very encouraging, given the independent approaches taken by the two packages to simulating the PSF. In particular, as the telescope begins taking data, having two methods to compare the PSF will greatly help ascertain the true optical model, while accurately accounting for various perturbations and defects. 

\section{Applications}
\label{sec:Additional}

\subsection{Filter reflection ghost}
Optical ghosts are a known effect in imaging surveys like Roman, and understanding them is essential for creating a comprehensive understanding of the Roman optical system. These effects are unavoidable, and are known to cause significant systematic errors for low surface-brightness science cases in particular \citep{Borlaff2022, Rodeghiero26, Pai26}. We include a physically motivated model for filter reflection ghosts in PSFSim to enable the community to simulate the impact of optical ghosts in their science cases and to develop tools for their correction and removal.

\begin{figure}
    \centering
    \includegraphics[width=\linewidth]{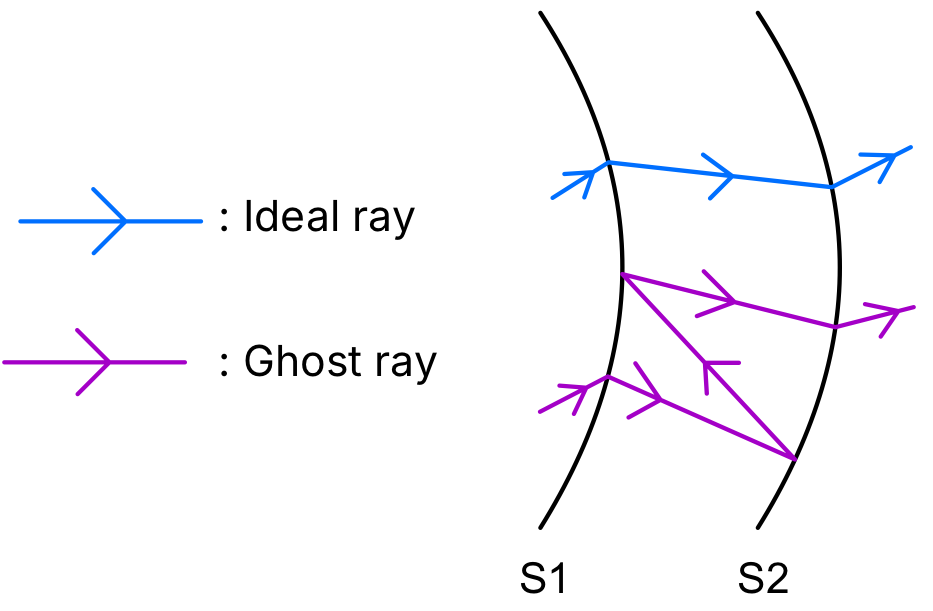}
    \caption{An illustration of an ideal ray (blue) and a ghost ray (purple) passing though the first and second surfaces (S1 and S2, respectively) of an imaging filter.}
    \label{fig:ghostpath}
\end{figure}

Before reaching the WFI, light passes through the Element Wheel, which contains 8 imaging filters, a prism, a grism, and a dark for calibration. Light travelling through an imaging filter will be partially transmitted and partially reflected. Ideal rays are transmitted through both surfaces (S1 and S2) of the filter. Some rays instead reflect at S2, reflect again at S1, and then are transmitted through S2. These internal reflections lead to displaced unfocused light appearing in images, referred to as an optical ``ghost'' (see Figure \ref{fig:ghostpath} for illustration). We refer to this particular optical ghost as the filter reflection ghost.

\begin{figure*}
    \centering
    \includegraphics[width=1.\linewidth]{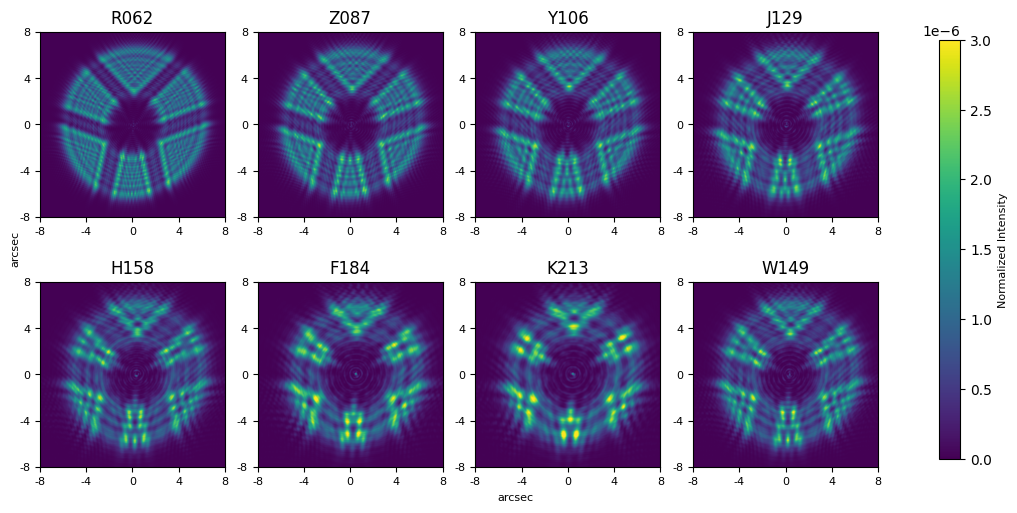}
    \caption{The Optical PSF for filter reflection ghosts across Roman imaging filters. The PSFs shown here are modelled at the center of SCA 9.}
    \label{fig:ghostsall}
\end{figure*}

The filter reflection ghost is modelled in PSFSim through the optional addition of the two filter internal reflections in the ray tracing path. The PSFSim user can initiate a ray bundle that includes the additional internal reflections to simulate optical ghosts by including the keyword \texttt{ghost=True} when initiating a \texttt{PSFObject}. For example, modifying the previous example given in Section \ref{sec:psf}:
\begin{lstlisting}[language=python]
ghostpsf = PSFObject(9,0,0, wavelength = 1.0, ghost = True)
ghostpsf.get_optical_psf()
plt.imshow(ghostpsf.Optical_PSF)
\end{lstlisting}
The results for all eight optical filters is shown in Fig.~\ref{fig:ghostsall}.
We see that for longer wavelengths, the diffraction features get more significant and extend farther from the shadows of the struts. 

%\elle{should we be worried about the ghostsall fig compiling in the conclusion section instead of with the other ghost stuff? when i try [H] it disappears}
%\kl{Nope dont worry about it-- the journal can always mess with formatting if they don't like it.}

\begin{figure*}
    \centering
    \includegraphics[width=\linewidth]{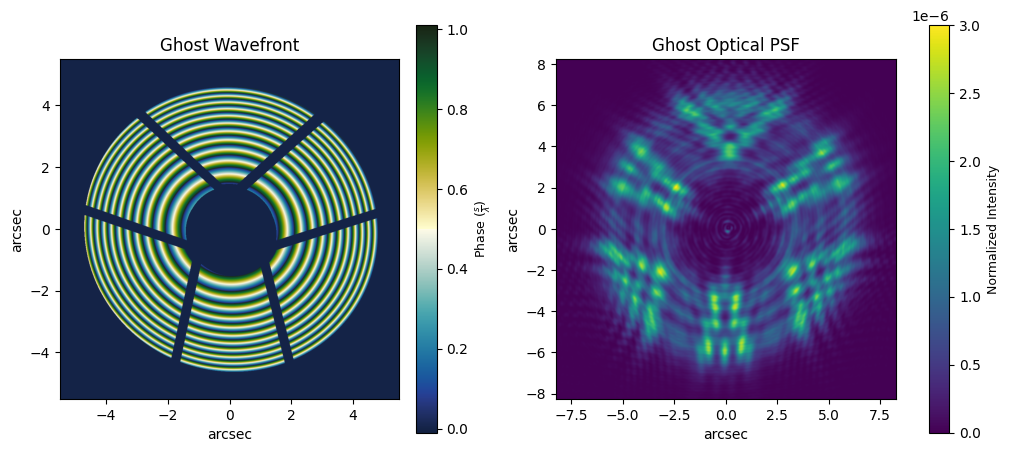}
    \caption{Left: Entrance pupil image of the simulated filter reflection ghost showing the wavefront phase in cycles. Right: The optical PSF for a filter reflection ghost. Both images are for a monochromatic $\lambda=1.58\, \mu$m ray bundle observed in the H band at the center of SCA 9.}
    \label{fig:sidebyside}
\end{figure*}

Figure \ref{fig:sidebyside} shows more detail on the filter refraction ghost. The left panel displays the wavefront of the ghost at the entrance pupil. The gaps from the secondary mirror and struts are clearly visible, and the wavefront phase oscillates more rapidly with increasing distance from the center. The PSF (the Fourier transform of the wavefront) shows the expected features; concentric rings are visible from the annular aperture and bright diffraction features appear around the struts. We also note the presence of an Arago spot in the center of the PSF image from Fresnel diffraction at the secondary mirror.

PSFSim does not model interference between the direct path and the filter reflection ghost. Since the filter reflection has $\sim 3$ cm of extra path length, in broadband light (as expected for most astronomical sources) any interference between the two paths would be washed out.

% \elle{should i put this text somewhere else? does the sidebyside fig need more description/discussion?}
%\kl{Added some more discussion, we should pass this by Chris}

% that impacts the optical component of the Roman PSF, and so they are taken into account for this physics first simulation tool. Their impacts span across surveys and focus areas, and so including filter reflection ghosts in our simulation enables us and the overarching science community to better prepare to model filter reflection ghosts, assess their effects on the data we collect as well as the science we perform, and correct for them.

% \elle{physics first model of psf... ghosts are physical optical effect
% will impact images
% so include in sim
% known effect in surveys, can help prepare for modelling and how to use for own purposes, eventual removal}

% \begin{itemize}
%     \item 1 paragraph + figure? on the filter ghost path and what we did to the code
%     \item 1 figure showing a wavefront plot (left panel, with rings) and a monochromatic ghost (right panel)
%     \item 1 paragraph on what we plan to do about ghosts for the WL analysis / how we want to use the ghost capability
% \end{itemize}
\subsection{Using diffraction spikes to calibrate optical aberrations}

The 12 diffraction spikes of the Roman Space Telescope can be used to calibrate the optical aberrations of the instrument. This was demonstrated for the Euclid space telescope \citep{2025A&A...697A...1E} by \citet{2026arXiv260625625N}, where the authors made use of the area of intersection between the diffraction spikes to relate to the defocus of the instrument given by the fourth order Zernike coefficient. The authors also demonstrate that a similar method can be adopted for Roman, which has even more diffraction spikes, and could therefore result in an even more precise measurement of the wavefront error. 

\begin{figure*}
    \centering
    \includegraphics[width=\linewidth]{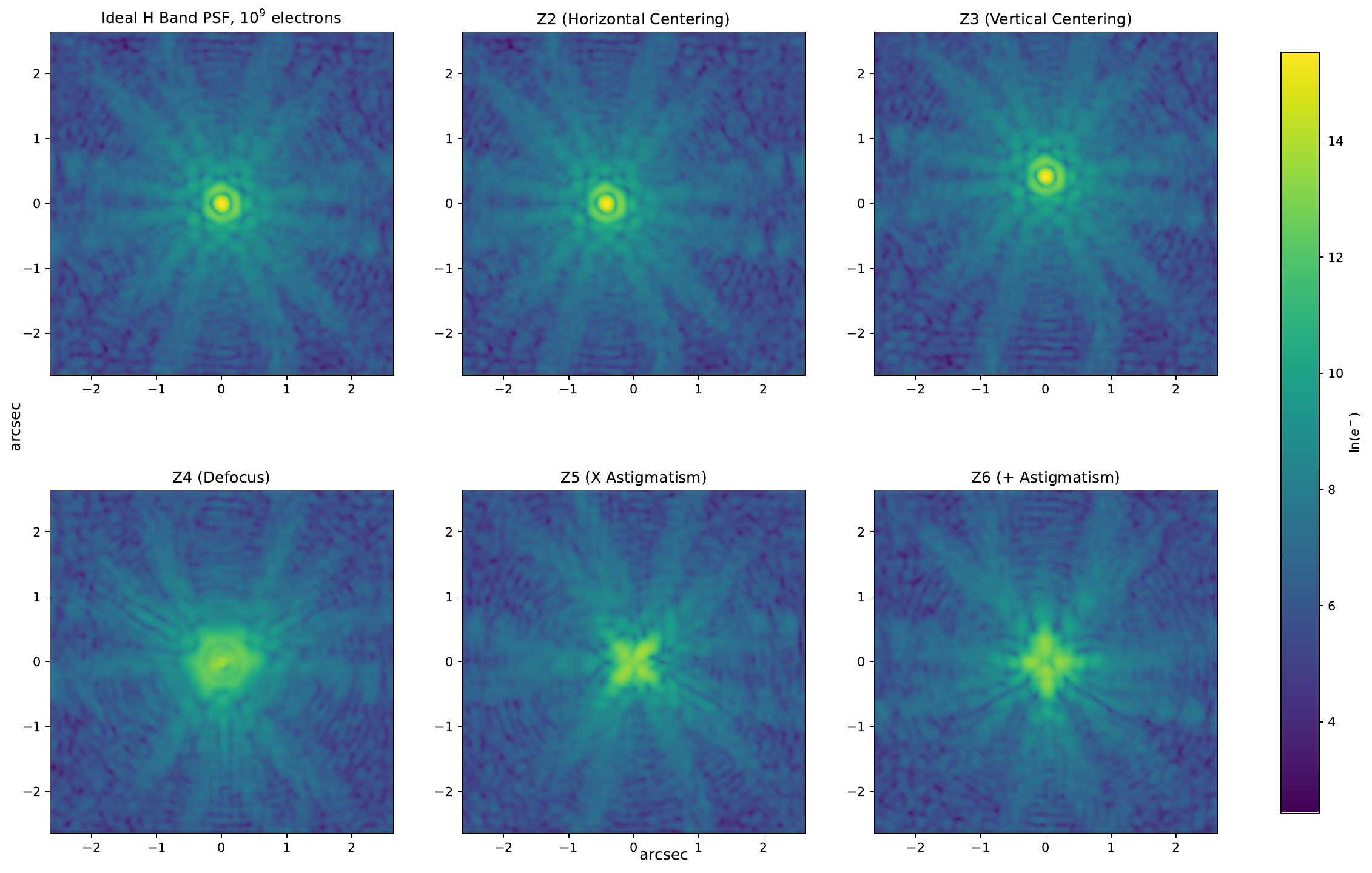}
    \caption{The H-band polyhcromatic PSF as it changes with the addition of various different exaggerated aberrations. We note the changes in the angles of the diffraction spikes of the PSF as the focus and astigmatism change.}
    \label{fig:diffspikes}
\end{figure*}

PSFSim has the ability to add in extra aberrations with the \verb|extra_aberrations| argument, where one can specify an array of additional zernike coefficients from $Z_2$ to $Z_6$. These are added to the optical path difference, and lead to visible shifts in the diffraction spikes, as displayed in Figure \ref{fig:diffspikes}.

Work by Kuhtenia et al. (in prep) suggests that the extra diffraction spikes in Roman images can be used to retrieve accurate $n=2$ (focus and astigmatism) wavefront errors from images of bright stars. The physical modelling for this effort will rely on PSFSim.

%find informative and precise priors for the wavefront errors in the telescope beyond just the focus term. In particular, using PSFSim, we can build an accurate forward model for the expected diffraction spikes, and back out accurate

\subsection{Polarization Effects}

Light from distant stars and galaxies can be linearly polarized in the optical and NIR wavelengths \citep{Fendt1996, Scarrott87, Jones00}, largely due to anisotropic scattering from spherical dust grains and selective extinction from aligned dust particles. It is thereby important to characterize the polarization dependence of the optical model \citep{Lin2020} which PSFSim takes into account.

In addition to correctly propagating the polarization as the light rays travel through the telescope, we also model the mirror and its effects on polarization.

Following the \citet{Born:1999ory} treatment on optical admittance through a stratified medium, we model our mirror surface coatings as a thin dielectric coating on silver. For the dielectric function, we used the model by \citet{Yang2015}. To approximate the mirror specifications released by the manufacturer, L3 Harris, and produce the expected zero-crossing of the linear retardance at $\sim 600$nm, we used a single layer thin film with index of refraction and thickness chosen to match the measured polarizations at 45 degree angle incidence. %used to say S and P polarizations

The major ``missing'' ingredient in this current model is the transmission coefficient (including wavelength and polarization dependence) of the filter coating. We plan to include a model for this in a future release of PSFSim.

We present the difference between this polarization dependent mirror model and an ideal mirror in Figure \ref{fig:polarization} below. We note that the difference between the two results in a small dipolar structure due to the short delay in propagation from light rays with different polarizations. 

We also perform adaptive ellipticity \citep{Hirata2003, Mandelbaum2005} measurements using the {\tt galsim.hsm} module \citep{2015A&C....10..121R} on the images across the focal plane, and note that the mean centroid difference is roughly 0.01 pixels in the $y$ direction, and the differences in $e_1$ between $(-7, -3) \times 10^{-4}$ and $e_2$ between $(-6, 6) \times 10^{-4}$. Here, $e_1$ and $e_2$ refer to the standard shear definition of ellipticity used for weak lensing measurements, and are in pixel coordinates. The negative $e_1$ values are reasonable, as the primary effect is along the $y$ axis, which corresponds to a negative $e_1$ in pixel coordinates.

\begin{figure*}
    \centering
    \includegraphics[width=\linewidth]{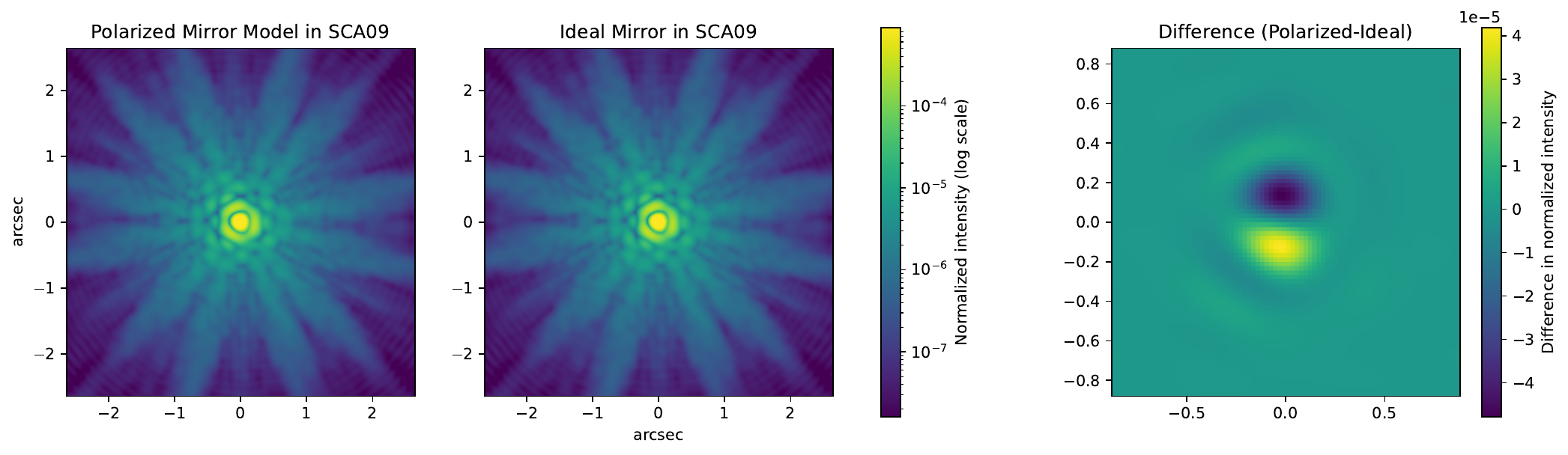}
    \caption{Left panel: the polychromatic PSF computed in H band for the centre of SCA09 assuming an imperfect mirror that treats polarizations differently. Middle panel: the polychromatic PSF computed in H band at the same location assuming an ideal mirror. Right panel: the difference between the two shows the expected dipolar structure. }
    \label{fig:polarization}
\end{figure*}

\section{Conclusion}

In this paper, we have presented PSFSim, a novel, physics-first simulation tool for modelling the Point Spread Function of the Nancy Grace Roman Space Telescope's Wide Field Instrument (WFI). By combining full optical ray tracing and incorporating field-dependent and wavelength-dependent aberrations, complex pupil geometry, and polarization with detailed physical models of detector level effects (such as anti-reflection coating interference, charge diffusion, and interpixel capacitance), PSFSim provides an end-to-end framework for generating realistic Roman PSFs and calibrating the optical model for Roman.

Key capabilities and findings of this work include:

\begin{itemize}
    \item \textbf{Physics Forward Architecture:} Accurate propagation of polarized electromagnetic fields through the telescope's optical train and directly into the active HgCdTe detector layers.
    \item \textbf{Flexible Perturbations:} The ability to introduce optical misalignments and environmental defects (such as ice buildup), allowing users to stress-test data reduction pipelines against non-ideal observatory states.
    \item \textbf{Validation \& Cross-Checks:} Successful comparison against existing tools like STPSF, showing strong overall agreement while highlighting critical subtle differences resulting from refraction and detector-level charge diffusion models.
\end{itemize}

As the Nancy Grace Roman Space Telescope begins science operations, high-precision PSF modelling will be foundational to achieving its core scientific objectives, from galaxy shape measurements in weak lensing to crowded-field photometry in microlensing and time-domain supernova cosmology. We will continue to update PSFSim based on in-flight measurements of the PSF, distortion, and throughput. PSFSim is fully open-source and publicly available to the astronomical community, offering a robust platform for validating project infrastructure pipelines, developing PSF reconstruction algorithms, and simulating performance under realistic operational conditions.

%Future updates to PSFSim will incorporate detector characterization data, updated models of internal optical ghosts, and refined optical and ice models as commissioning concludes.

\section*{Acknowledgements}
The authors were supported by NASA ROSES grant 22-ROMAN11-0011 via a JPL subaward. We thank Federico Berlfein, Mike Jarvis, Marcio Melendez Hernandez, and Marhsall Perrin for useful discussions. Computations were performed on the Ohio Supercomputer Center \citep{OSC}.

\end{document}